\documentstyle[aaspp4,12pt,psfig]{article}
\begin{document}

\newcommand{\ltsim}{\lower.5ex\hbox{$\; \buildrel < \over \sim \;$}}
\newcommand{\gtsim}{\lower.5ex\hbox{$\; \buildrel > \over \sim \;$}}
\newcommand{\order}[1]{\mbox{$\cal{O}$ ({#1})}}
\newcommand{\etal}{\mbox{\it et~al.}}
\newcommand{\suph}{\mbox{$^{\rm h}$}}
\newcommand{\supm}{\mbox{$^{\rm m}$}}
\newcommand{\sups}{\mbox{$^{\rm s}$}}

\def\arcmin{\hbox{$^\prime$}}
\def\arcsec{\hbox{$^{\prime\prime}$}}
\def\degs       {$^\circ$}
\def\grs        {$\gamma$-rays }
\def\gr         {$\gamma$-ray }
\def\pers       {$\rm s^{-1}$}
\def\ergps      {$\rm erg$ ~\pers }
\def\etal     {$et~al.$}
\def\apr{$\sim$}
\def\asca {$ASCA\,$}
\def\fluxunit {erg cm$^{-2}$s$^{-1}$kev$^{-1}$}
\def\etal {{\em et al.}}
\def\spose#1{\hbox to 0pt{#1\hss}}
\newcommand\lsim{\mathrel{\spose{\lower 3pt\hbox{$\mathchar"218$}}
     \raise 2.0pt\hbox{$\mathchar"13C$}}}
\newcommand\gsim{\mathrel{\spose{\lower 3pt\hbox{$\mathchar"218$}}
     \raise 2.0pt\hbox{$\mathchar"13E$}}}

\newcommand{\units}[1]{\mbox{$\rm\,#1$}}
\newcommand{\kpc}       {\mbox{\rm\,kpc}}
\newcommand{\Hz}        {\mbox{\rm\,Hz}}
\newcommand{\kHz}       {\mbox{\rm\,kHz}}
\newcommand{\MHz}       {\mbox{\rm\,MHz}}
\newcommand{\GHz}       {\mbox{\rm\,GHz}}
\newcommand{\erg}       {\mbox{\rm\,erg}}
\newcommand{\Jy}        {\mbox{\rm\,Jy}}
\newcommand{\mJy}       {\mbox{\rm\,mJy}}
\newcommand{\uJy}       {\mbox{$\,\mu\rm{Jy}$}}
\newcommand{\K}         {\mbox{\rm\,K}}
\newcommand{\KperJy}    {\mbox{$\rm\,K\,Jy^{-1}$}}
\newcommand{\dm}        {\mbox{$\rm\,pc\,cm^{-3}$}}
\newcommand{\kms}       {\mbox{$\rm\,km\,s^{-1}$}}
\newcommand{\yr}        {\mbox{\rm\,y}}
\newcommand{\s}         {\mbox{\rm\,s}}
\newcommand{\ms}        {\mbox{\rm\,ms}}
\newcommand{\us}        {\mbox{$\,\mu\rm{s}$}}
\newcommand{\cm}         {\mbox{\rm\,cm}}
\newcommand{\m}         {\mbox{\rm\,m}}
\newcommand{\degyr}     {\mbox{$\,^\circ\,{\rm y}^{-1}$}}
\newcommand{\degree}    {\mbox{$^\circ$}}
\newcommand{\gauss}     {\mbox{\rm\,G}}
\newcommand{\accel}     {\mbox{$\rm\,m\,s^{-2}$}}
\newcommand{\Msun}      {\mbox{$\,M_{\mathord\odot}$}}
\newcommand{\Lsun}      {\mbox{$\,L_{\mathord\odot}$}}
\newcommand{\Rsun}      {\mbox{$\,R_{\mathord\odot}$}}

\newcommand{\lsi}{\mbox{LSI\,+61\degree\,303}}
\newcommand{\eg}{\mbox{\it e.g.~}}
\newcommand{\gt}{\mbox{GT~0236+610}}

\title{Simultaneous X-ray and Radio Monitoring of the Unusual Binary
\lsi: Measurements of the Lightcurve and High-Energy Spectrum}
\author{F. A. Harrison\altaffilmark{1}}
\affil{Space Radiation Laboratory, California Institute of Technology,
	Pasadena, CA 91125}
\author{P. S. Ray\altaffilmark{2}}
\affil{ Code 7621 
	Space Science Division,
	Naval Research Laboratory,
	Washington, DC~~20375}
\author{D. A. Leahy\altaffilmark{3}}
\affil{Department of Physics,
	University of Calgary,
	Calgary, Canada T2N~1N4}
\author{E. B. Waltman\altaffilmark{4}}
\affil{ Code 7210 
	Remote Sensing Division,
	Naval Research Laboratory,
	Washington, DC~~20375}
\author{G. G. Pooley\altaffilmark{5}}
\affil{Mullard Radio Astronomy Observatory, Cavendish Laboratory,
       Madingley Road, Cambridge CB3 0HE}

\altaffiltext{1}{E-mail: fiona@srl.caltech.edu}
\altaffiltext{2}{E-mail: paulr@xeus.nrl.navy.mil}
\altaffiltext{3}{E-mail: leahy@iras.ucalgary.ca}
\altaffiltext{4}{E-mail: waltmane@rsd.nrl.navy.mil}
\altaffiltext{5}{E-mail: ggp1@mrao.cam.ac.uk}

\begin{abstract}

The binary system, \lsi , is unusual both because of the dramatic,
periodic, radio outbursts, and because of its possible association
with the 100~MeV gamma-ray source, 2CG~135+01.  We have performed
simultaneous radio and {\em Rossi X-ray Timing Explorer} X-ray
observations at eleven intervals over the 26.5 day orbit, and
in addition searched for variability on timescales ranging from
milliseconds to hours.  We confirm the modulation of the X-ray
emission on orbital timescales originally reported by Taylor {\em et
al.} (1996), and in addition we find a significant offset between the
peak of the X-ray and radio flux.  We argue that based on these
results, the most likely X-ray emission mechanism is inverse
Compton scattering of stellar photons off of electrons accelerated at
the shock boundary between the relativistic wind of a young pulsar and
the Be star wind.  In these observations we also detected 2 -- 150 keV
flux from the nearby low-redshift quasar QSO~0241+622. Comparing
these measurements to previous hard X-ray and gamma-ray observations
of the region containing both \lsi\ and QSO~0241+622,  it is clear
that emission from the QSO dominates.
\end{abstract}

\keywords{radio continuum: stars --- stars: individual (\lsi) ---
stars: binaries --- X-rays: binaries}

\section{Introduction}

The Be binary system \lsi\ (associated with the radio source GT
0236+610) is remarkable for dramatic radio outbursts occurring with
the 26.5 day orbital cycle (\cite{gt78,tg82}), and for its possible
association with the 100 MeV gamma-ray source 2CG~135+01
(\cite{bh83,kab+97}).  Radio flares lasting several days occur every
orbit, with the peak flux varying in phase by up to half the orbital
cycle.  The binary is a weak, variable X-ray source
(\cite{gm95,typ+96,lhy97}) with a non-thermal spectrum.  EGRET
measurements indicate that the 100~MeV emission from the region may
also be variable~(\cite{tkm+98}).  The gamma-ray error box contains no
other likely counterparts.  The gamma-ray localization is not,
however, sufficiently accurate to make a secure association between
2CG~135+01 and \lsi .

The mechanisms responsible for the radio outbursts and variable X-ray
emission are not well understood. The radio emission is consistent
with optically-thin synchrotron radiation for most of the outburst,
with indication that the source is self-absorbed at the beginning of
the outburst rise (\cite{tg84}). The X-ray emission is consistent with
either an extension of the radio synchrotron spectrum, or inverse
Compton scattering of the stellar photons by the relativistic
electrons responsible for the radio flux (\cite{lhy97}).  The most
plausible models proposed for the origin of the relativistic electrons
are acceleration in a shock produced at the interaction of the Be star
wind with the relativistic wind of a young pulsar companion
(\cite{mt81}), or in a shock produced by the ejection of matter
accreted onto a pulsar magnetosphere (\cite{csmc95}).  In either case,
both the X-ray and radio emission would vary with orbital phase.

In this paper, we present simultaneous {\em RXTE} X-ray, Green Bank
Interferometer (GBI) and Ryle Telescope radio observations taken over
a single orbital cycle.  These observations provide a complete orbital
light curve in 2--10 keV X-rays and 2.25 GHz, 8.3 GHz, and 15 GHz
radio frequencies.  On only one prior occasion has \lsi\ been
monitored in the X-ray band over an entire orbital cycle
(\cite{typ+96}).  We have also analyzed archival X-ray data and
compiled a history of the X-ray variability.  We have established
X-ray flux variations on hourly timescales, and have performed the
most sensitive searches to date for rapid X-ray pulsations from the
suggested neutron star companion.  In addition, we have analyzed the 2
-- 150 keV spectra of both \lsi\ and the nearby quasar QSO~0241+622,
which has been included in the field of view of many previous hard
X-ray measurements.

\section{Orbital Variability}

\subsection{RXTE Observations}
\label{sec-obs}

\begin{table}
\caption{Observation log of RXTE observations of \lsi, with fitted
fluxes.}
\label{tab-obs}
\begin{tabular}{cccccccc}
\hline\hline
Date & Radio Phase & \# PCU & PCA           & Flux (2--10 keV) \\
UTC &              &        & live time (ks)& $\times 10^{-12}$ erg cm$^{-2}$ s$^{-1}$ \\
\hline
1996 Mar 01 13:57--19:25 & -0.17 & 5 & 8.9 & 8.4 \\
1996 Mar 04 21:38--01:36 & -0.04 & 5 & 9.3 & 12.1 \\
1996 Mar 07 23:33--03:28 & 0.07 & 5 & 9.4 & 6.9 \\
1996 Mar 10 01:07--05:02 & 0.15 & 5 & 9.1 & 6.5 \\
1996 Mar 13 04:06--08:19 & 0.27 & 5 & 10.0 & 9.6 \\
1996 Mar 16 00:54--06:06 & 0.38 & 5 & 9.5 & 13.2 \\
1996 Mar 18 10:44--15:48 & 0.47   & 5 & 9.5 & 20.0 \\
1996 Mar 20 23:19--03:06 & 0.58 & 0 & 0 & NA \\
1996 Mar 24 07:55--13:10 & 0.69 & 3 & 11.0 & 12.2 \\
1996 Mar 26 00:56--06:46 & 0.75 & 5 & 13.7 & 10.3\\
1996 Mar 30 04:07--10:40 & 0.91 & 5 & 14.6 & 9.2\\
\hline
\end{tabular}
\end{table}

NASA's {\em Rossi X-ray Timing Explorer} ({\em RXTE}) carries two
pointed instruments, the Proportional Counter Array (PCA) and the
High-Energy X-ray Timing Experiment (HEXTE).  The PCA is an array of 5
nearly identical xenon proportional counter units (PCUs) sensitive to
X-rays in the range 2--60 keV.  The PCUs are co-aligned and collimated
to a 1\degree\ field of view (FWHM).  The total collecting area of the
PCA is $\sim 6300 \cm^2$. The HEXTE consists of two clusters of
NaI/CsI phoswich scintillation counters sensitive in the range 15--250
keV.  Each cluster has an effective area of $\sim 800 \cm^2$.  The
clusters rock such that one is always pointed at the source, while the
other monitors the background at a slightly offset point.  The HEXTE
collimators also have a field of view of 1\degree (FWHM).

We observed \lsi\ eleven times with the PCA and HEXTE between 1996
March 1 and 1996 March 30, at intervals approximately equally spaced
over the 26-day orbit.  Several problems with the instruments affected
our observations (which were made only 1 month after guest
observations with RXTE began).  On 1996 March 6, one of the four
detectors in HEXTE cluster B failed.  On 1996 March 19, PCUs 4 and 5
exhibited anomalous behavior, resulting in all 5 of the PCUs being
shut down for one of our observations, and only 3 PCUs being active
for the next.  On 1996 March 25, the PCA gain was changed. The
observation times are shown in Table~\ref{tab-obs}, along with the
radio phase, on-source integration time (total time minus SAA passages
and Earth occultations), and number of active detectors.  

Most previous papers on this source have defined the radio phase
according to the ephemeris of Taylor and Gregory (1984)\nocite{tg84}.
They defined an arbitrary phase zero of JD 2443366.775 and a period
measured to be $26.496 \pm 0.008$ days based on data taken between
1977 and 1982.  Using this ephemeris, the peak of the radio outbursts
typically occurs around phase 0.6.  More recently, Ray \etal\
(1997)\nocite{rfw+97} found that radio outbursts monitored from
1994--1996 required using a period of $26.69 \pm 0.02$ days.  They
suggested that the orbital phase of the radio outbursts was varying in
response to some other parameter in the system.  In a Bayesian
analysis of the complete 20-year data set Gregory, Peracaula and
Taylor (1999)\nocite{gpt99} find that the most favorable model for
these data is a constant orbital period of $26.4917 \pm 0.0025$ d with
a longer period of $1584^{+14}_{-11}$ d which modulates the typical
outburst phase with a magnitude of 8 days and the flux between 100 and
300 mJy.  In this paper we quote radio phase referring to this orbital
ephemeris ($T_0 = $JD 2443366.775, $P_{\rm orb}$ = 26.4917 d).  This
allows a consistent treatment of current and historical observations.
When appropriate, we will quote the predicted outburst phase from the
long term modulation analysis above.

\subsection{RXTE Analysis and Results}

\subsubsection{PCA}

We analyzed the PCA observations using the standard FTOOLS 4.0 package
provided by the {\em RXTE} Guest Observer Facility (GOF).  For most
analyses not requiring high time resolution, we used the ``Standard
2'' mode data.  This mode accumulates a single 129 channel spectrum
from each detector anode every 16 seconds.  These data are filtered by
requiring that the PCUs be on, the source be more than 10\degree\
above the Earth's limb, and the spacecraft be pointed within
0.02\degree\ of the source.  Because the source is faint, and detected
only at low energies, signal to noise is improved by taking data only
from layer 1 of the PCUs.  We then combined all of the selected data
into one spectrum for each observation.  Finally, a background
spectrum is estimated and subtracted from the detected counts.

Because the PCA does not rock on and off source, to determine the
background counting rate we rely on a parametrized model of the
background, {\tt pcabackest v1.5} (\cite{jah+98}), provided by
the {\em RXTE} GOF.  This model includes 3 components: internal
background which is a function of the instantaneous veto rate measured
in the PCA, internal activation background which is a function of the
time since the last SAA passage, and the cosmic diffuse flux in the
1\degree\ aperture which is an average of high Galactic latitude
blank-sky pointings.

\begin{figure}
\centerline{\psfig{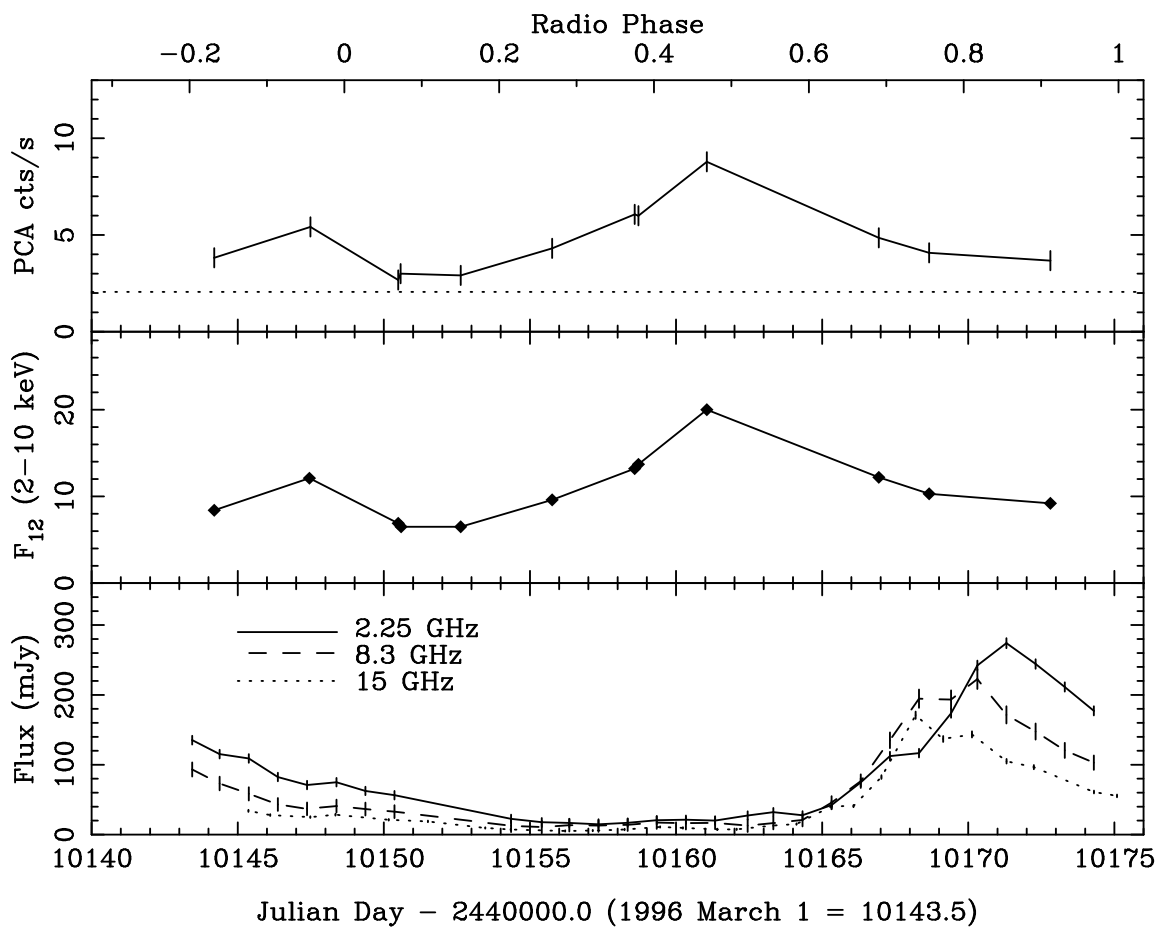}}
\caption[fig-xtelight.ps]{
Comparison of XTE PCA flux and radio flux during 1996
March. The top panel shows the PCA count rate (5 PCUs, 2--10 keV,
Layer 1 only) for each observation.  The dotted line shows the
2$\sigma$ upper limit to the error in the zero level due to errors in
the model of the diffuse X-ray flux in the aperture (see text). The
second panel shows the corresponding fitted 2--10 keV fluxes in units
of $10^{-12}$ erg cm$^{-2}$ s$^{-1}$.  The bottom panel shows the
radio lightcurves at 2.25, 8.3, and 15 GHz measured
quasi-simultaneously.  Radio phase is defined in the text.\label{fig-xtelight}}
\end{figure}

The top panel of Figure~\ref{fig-xtelight} shows the
background-subtracted countrate in the 2--10~keV band for the ten
observations with useful PCA data.  Source variability is evident,
with a peak countrate occurring between radio phase 0.45 -- 0.6.  The
second panel shows the X-ray flux, calculated from spectral fits to
these data.  Using XSPEC v10.00, we fit the individual pointings with
a power-law spectral model, with hydrogen column density fixed at the
value measured by {\em ASCA} of $0.6 \times 10^{22}$ cm$^{-2}$
(\cite{lhy97}).  The fluxes were calculated by integrating this
best-fit model from 2 to 10 keV.

The dominant error in the source variability and flux measurements is
uncertainty in the background. Errors in estimating the time-variable
internal background component will contribute to errors in determining
the source variability, while errors in determining the cosmic and
Galactic diffuse flux contributions, or other X-ray sources in the FOV
result in an uncertain zero level.  To estimate errors in internal
background determination, we produced background-subtracted countrate
spectra for Earth-occulted data taken during our observations.
Variations in the countrates are a measure of imperfect subtraction of
the internal background components.  We used the dispersion in these
values to derive the systematic error bars on the individual
observation countrates shown in Figure~\ref{fig-xtelight}. These
systematic errors dominate the statistical errors.

The horizontal dotted line in the top panel of Figure~\ref{fig-xtelight}
shows the 2$\sigma$ upper limit in the uncertainty in zero level in
the countrates due to the fact that the diffuse background may be
different in the direction of \lsi\ than in the fields used for
background subtraction.  Two factors contribute to this uncertainty:
variations of the diffuse cosmic background and any unaccounted for
contribution from the diffuse Galactic ridge emission.  The former
contributes an estimated 0.6 count/s (1 $\sigma$) uncertainty. This is
determined from scaling the fluctuations measured by {\em HEAO-1}
A-2 (\cite{mj92}), correcting by the square root of the ratio of the
{\em HEAO-1} A-2 to PCA beam sizes.  We derive an upper limit of 0.5
counts/s (1 $\sigma$) to any contribution from diffuse Galactic ridge
emission in two separate ways.  First, the upper limit from {\em
HEAO-1} A-2 presented in Worrall {\em et al.} (1982)~\nocite{wmb+82}
toward this direction gives an upper limit of 0.4 counts/s, which
compares well to the 0.5 counts/s upper limit derived from PCA
slew data taken from a 0.5 -- 2\degree\ annulus around \lsi .
 
Confusion from other X-ray sources in the FOV is not a problem in this
field.  The two sources found in the {\em ROSAT} sky survey within the
PCA field of view are both stellar sources with soft spectra which, based
on spectral fits, contribute $<$0.1 counts/s in the PCA.  One
of these sources, identified with the stellar source BD 60536,
was also detected with {\em ASCA}, which found a thermal spectrum
with temperature 0.8~keV (\cite{lhy97}).

\subsubsection{HEXTE}

\begin{figure}
\centerline{\psfig{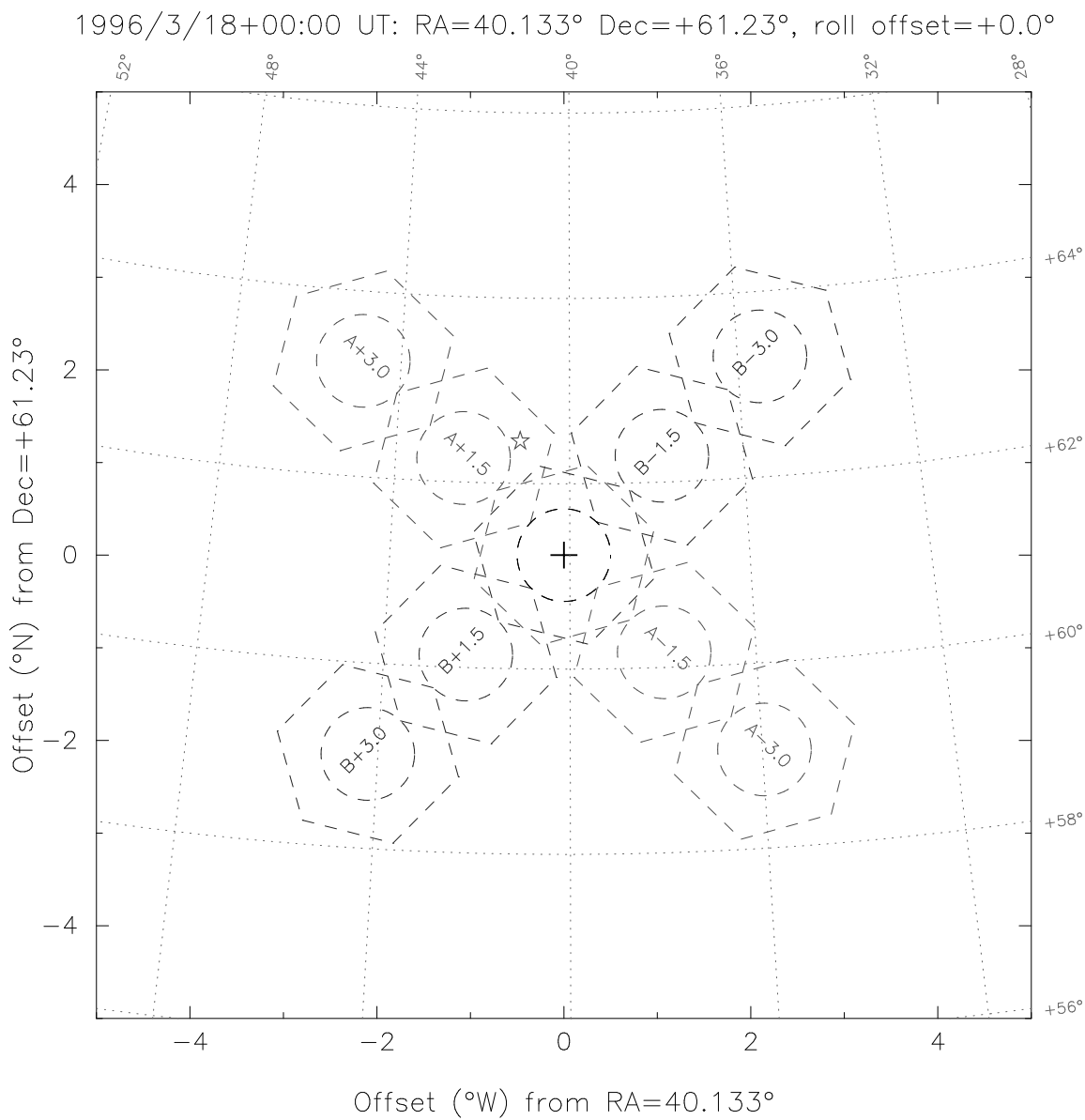}}
\caption[fig-hexteqso.ps]{Orientation of the HEXTE background fields
during observation 7. The star shows the position of the quasar
QSO~0241+662, which was contained in one of the background fields,
A+1.5.  The orientation of the fields varied between pointings.\label{fig-hexterock}
}
\end{figure}

During the observations the two HEXTE clusters were configured to
alternately point on-source and at four separate background regions
1.5 degrees away.  The HEXTE data analysis proceeded as follows.  We
first selected events by which field the cluster was viewing.  We then
extracted spectra for each field.  At this point, we corrected the
exposure times for each spectrum to account for the dead time in the
HEXTE detector using {\tt hxtdead v0.0.1}.  Due to electronics
problems, the dead time is much larger than was expected before
launch.

During six of the eleven observations, one of the background regions
contained the quasar QSO 0241+622 within the field of view of the
HEXTE collimator.  This QSO is a known hard X-ray source (\cite{tp89})
and many previous hard X-ray observations of the region made using
non-imaging instruments have been unable to conclusively determine
whether \lsi\ or the QSO is the source of the measured flux.
Figure~\ref{fig-hexterock} shows the four background locations,
labeled as A$\pm$1.5\degree\ and B$\pm$1.5\degree\ during pointing 7,
showing the position of QSO 0241+622 (indicated by a star) just
outside the FWHM FOV of background region A+1.5\degree.  By using one
of the other three fields that do not contain any known hard X-ray
source as background fields, we were able to derive flux measurements
both for \lsi\ and for QSO 0241+622.  We did not use the A+1.5\degree\
field as a background for the \lsi\ pointings.

\lsi\ is not bright enough in the HEXTE band to obtain significant
detections for the individual pointings.  We therefore summed data
from all pointings, totaling 69.8 ks on source after deadtime
correction, to search for high-energy emission.  \lsi\ is detected at
the $6.7 \sigma$ level in the 15--150 keV band in the summed data set
($0.45 \pm 0.1$ counts/s in Cluster A and $0.37 \pm 0.07$ counts/s in
Cluster B). We also divided the observations into two sets, set L
consisting of pointings 3, 4, 10, and 11, when the measured PCA flux
was low (below the mean), and set H consisting of pointings 6 -- 9
when the measured PCA flux was high.  In set L, we did not find a
significant detection of the source, with a 15--150 keV countrate of
$-0.014 \pm 0.15$ in Cluster A and $0.21 \pm 0.11$ c/s in Cluster B.
In set H, however, we measured a positive countrate in the 15--150 keV
band of $0.86 \pm 0.16$ in Cluster A and $0.49 \pm 0.11$ in Cluster B,
a detection at the $7\sigma$ level.  This gives further confidence
that the measured hard X-ray flux is associated with
\lsi , and is modulated on the same timescale as the soft emission.  
For set H, model fitting with XSPEC using a power law spectral model
yields a flux of $9 \times 10^{-11}$ erg cm$^{-2}$ s$^{-1}$ in the
15--150 keV band.

For QSO 0241+622, we summed data from pointings 6--11, during which
the quasar was in field A+1.5\degree\ and within 0.7\degree\ of the
field center.  Using field A--1.5\degree\ to determine the background,
we detect the quasar at a significance level of 5.3$\sigma$ in the
15--150 keV band, with a countrate in the instrument of $0.85 \pm
0.16$.  We also summed data from pointings 1--5, when the quasar was
further than 0.7\degree\ from the field center. Using the same
technique we find a countrate consistent with zero of $0.21 \pm 0.17$.
Correcting for the collimator response, and using XSPEC to fit a
power-law to pointings 6 -- 11, we find a hard X-ray flux for QSO
0241+622 of $2.0 \times 10^{-10}$ erg cm$^{-2}$ s$^{-2}$ (15--150
keV).  We find a power law photon spectral index of 1.2--2.2, but this
may be modified by the collimator somewhat.  Given the limited
statistics of the detection, applying an energy-dependent collimator
correction is not warranted.

To further confirm the detection of QSO 0241+622, we obtained a
10~ksec observation of the source taken 1997 October 21 from the
public archive, and analyzed the PCA data (the observation is too
short for a significant HEXTE detection).  QSO 0241+622 is positively
detected in this pointing, with a PCA 2--10 keV count rate of about 11
counts/s in 3 PCUs.  The background estimate for this period is much
better than for the 1996 March observations due to a recent background
model developed by the PCA team which is applicable to gain epoch 3.
Model fitting with an absorbed power law plus Gaussian iron line at
6.1 keV yields an excellent fit (reduced $\chi^2$ of 0.6 when fitting
Standard2 channels 3 to 38).  The best-fit power law spectral index is
$1.73 \pm 0.05$ with normalization of $0.009 \pm 0.001$
$\gamma$/cm$^2$/s/keV at 1~keV.  The corresponding 2--10 keV flux is
$3.6 \times 10^{-11}$~erg/cm$^2$/s. Extrapolating this fit into the
15--150 keV HEXTE band yields a flux of $1.0 \times 10^{-10}$
erg/cm$^2$/s, reasonably close to our off-axis HEXTE observation
considering the possibility of source variability, and the rough
collimator correction and poor statistics of our HEXTE measurement.
The measured flux and spectral index are fully consistent with the earlier
{\em EXOSAT} measurement of $3.7 \times 10^{-11}$ erg/cm$^2$/s (2--10
keV) and spectral index of $1.70^{+0.27}_{-0.18}$ (\cite{tp89}).

\subsection{Radio Data Analysis and Results}

Throughout the entire binary orbit coincident with the {\em RXTE}
pointings, we monitored the source frequently using the two-element
Green Bank Interferometer (GBI) in West Virginia and the Ryle telescope
in Cambridge, England.  The GBI consists of two 26 meter antennas each
of which has a pair of cooled receivers which simultaneously receive
signals at 2.25 and 8.3 GHz with a system bandwidth of 35 MHz.  Each
observation consists of a ten minute scan. Measured correlator
amplitudes are converted to fluxes by comparing to standard, regularly
observed calibrators.  Constraints imposed by the radio telescope
mount and elevation angle of the source precluded strict simultaneity
of the X-ray and radio observations, however radio observations were
performed closely prior to and following the {\em RXTE} pointings.
Details of the data analysis and error estimation for GBI monitoring
of \lsi\ were presented by Ray \etal (1997)\nocite{rfw+97}.

The Ryle Telescope is an 8-element radio interferometer.  Each element
is a 13-m Cassegrain antenna.  Throughout the observations described
here, the instrument operated at 15 GHz with a bandwidth of 350 MHz.
The RMS noise on a 5-minute integration is about 1.7 mJy, although
this degrades in poor weather. When the source is bright, the
overall flux calibration will be the dominant error; and RMS deviation
of 3\% has been established empirically for these circumstances.
Observations of a calibration source (4C67.05) are interleaved with
observations of \lsi.

The bottom panel of Figure~\ref{fig-xtelight} shows the three-frequency
radio data for the orbital cycle coincident with the {\em RXTE}
pointings.  As seen previously by Ray \etal (1997)\nocite{rfw+97},
these data show a clear progression from an approximately flat
spectrum to an optically-thin spectrum during the outburst.

\subsection{Archival Data}

\lsi\ has been simultaneously monitored at both X-ray and radio
wavelengths on two other occasions, once with ROSAT/VLA (\cite{typ+96}),
and once with ASCA/GBI (\cite{lhy97}).  In order to compare the
variability seen with XTE, we have reanalyzed these data, using
the updated ephemeris. In the case of ROSAT, we reduced the
X-ray data using the accurate column depth measured by ASCA and a power law
model to derive the flux.  The GBI data taken simultaneously with
the ASCA measurements have not been previously presented.

\subsubsection{ASCA}

\begin{figure}
\centerline{\psfig{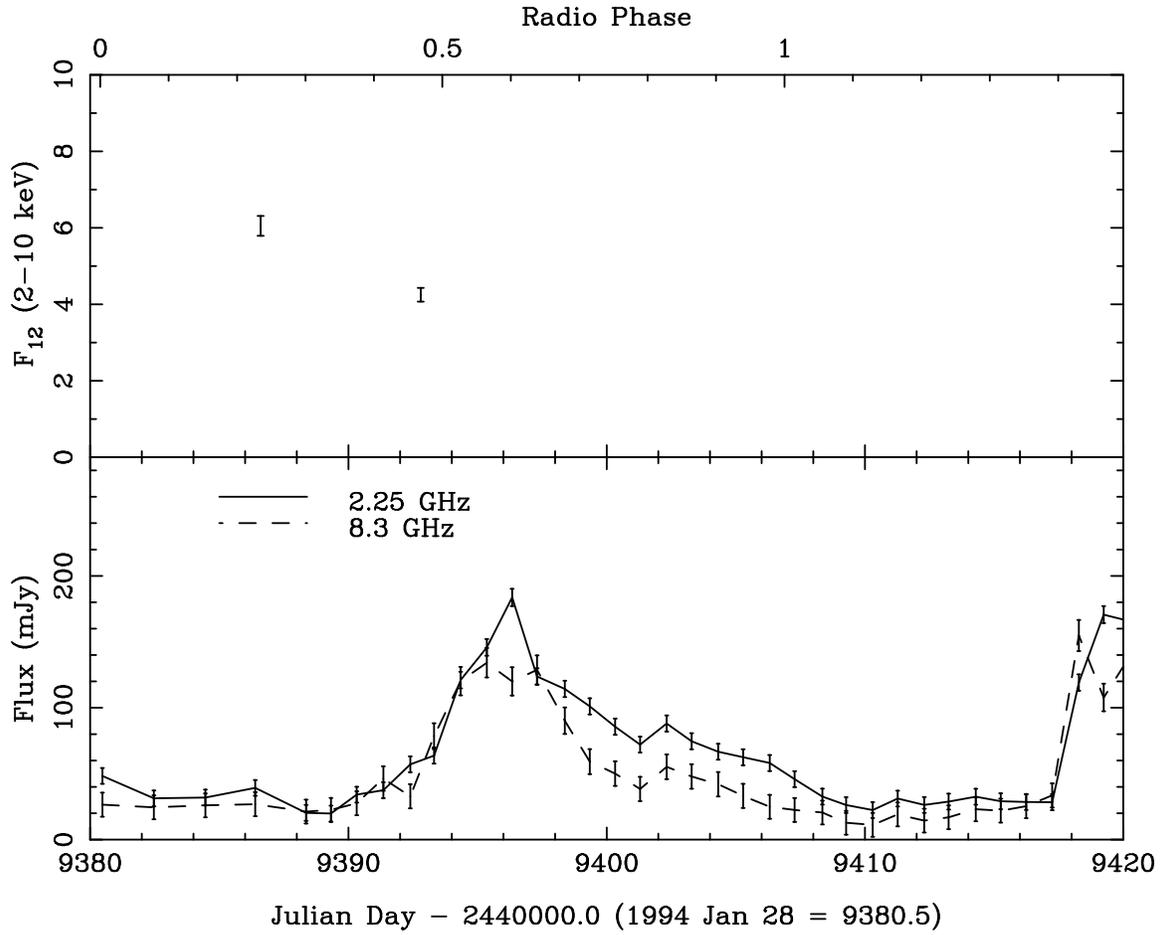}}
\caption[fig-ascalight.ps]{The two-frequency
radio lightcurve(bottom panel) taken simultaneously
with the two ASCA pointings (flux levels shown in the top
panel). All error bars are one sigma. \label{fig-ascalight}
}
\end{figure}

The source was observed twice during the same orbit by \asca . The
first observation took place on 1994 February 3 (MJD 49386) beginning
at 7:34 UT and ending at 11:01 UT, and the second observation began on
1994 February 9 (MJD 49392) at 7:34 UT and ended on the same day at
17:08 UT.  Throughout the radio outburst coincident with the \asca\
pointings, we monitored the source using the Green Bank Interferometer
(GBI) at 2.25 GHz and 8.3 GHz. During the quiescent radio state \lsi\
was monitored every two days, and during the radio outburst the source
was monitored 2 -- 4 times every day.  Each observation consists of a
ten minute scan, during which both frequencies are monitored
simultaneously.

Using the updated ephemeris, the ASCA pointings occurred at phases
0.23 and 0.47.  Figure~\ref{fig-ascalight} shows the X-ray fluxes
for the two pointings plotted with the radio data taken during
the same outburst.  Derivation of the flux values was presented
by Leahy, Harrison and Yoshida (1997).

\subsubsection{ROSAT}

\begin{figure}
\centerline{\psfig{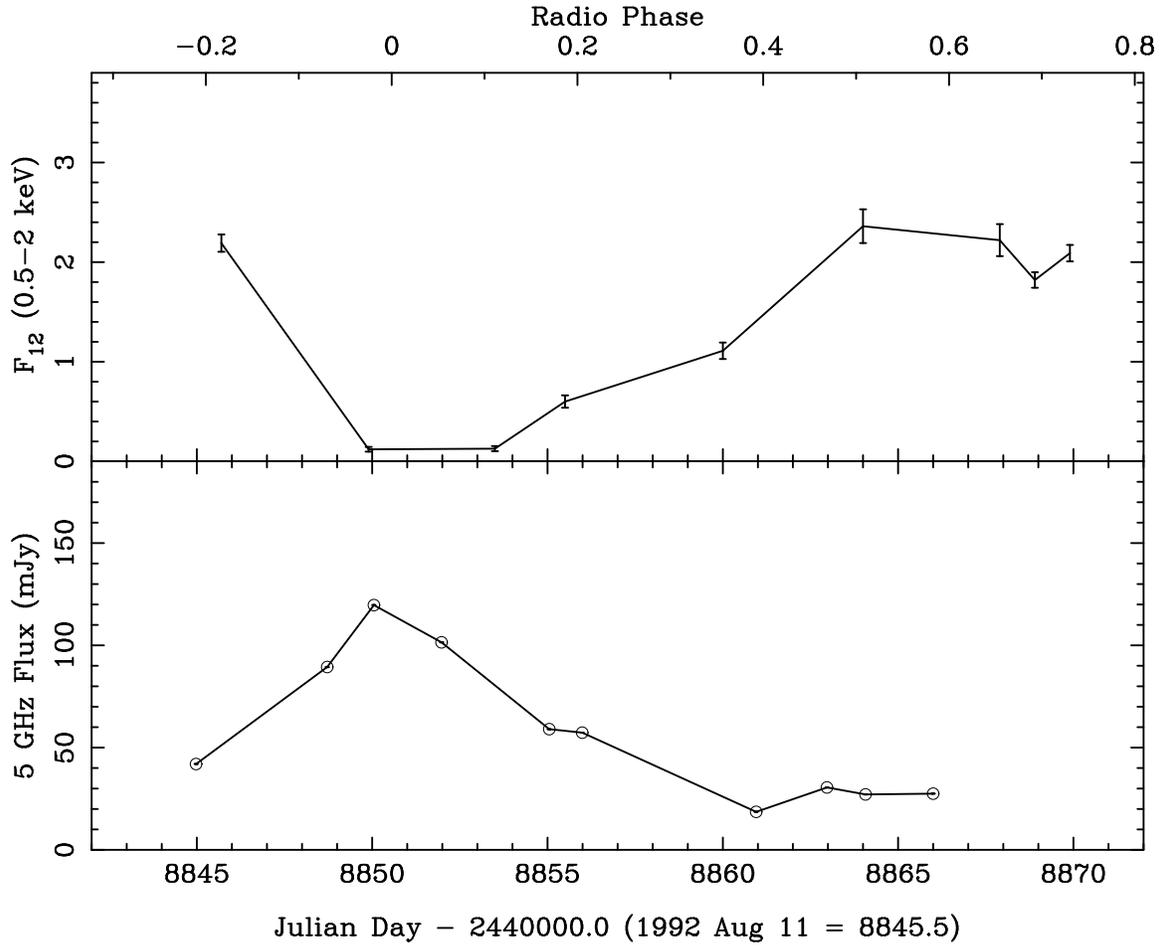}}
\caption[fig-rosatlight.ps]{The ROSAT light curve from one radio
outburst cycle (top panel) and the 5 GHz radio data taken at the VLA during
the same cycle (bottom panel).\label{fig-rosatlight}
}
\end{figure}

Taylor {\em et al.} (1996) observed \lsi\ on nine occasions over an
orbital cycle with ROSAT, and also with the VLA at 5 GHz on ten occasions
during the same outburst.  We have replotted their radio data,
correcting for the updated ephemeris, in Figure~\ref{fig-rosatlight}.
The X-ray flux points in this plot are based on a reanalysis of data
retrieved from the public archive. From the ASCA observations, it is
now clear that the spectrum is a power law, and an accurate column
depth has been derived.  For fitting the X-ray data, we allowed the
powerlaw spectral index to vary, but fixed the column density at the
ASCA value of $0.6 \times 10^{22}$ cm$^{-2}$ (\cite{lhy97}).  The
X-ray flux is also plotted in Figure~\ref{fig-rosatlight}.

\section{Short-term Variability}
\label{sec-aperiodic}

\begin{figure}
\centerline{\psfig{file=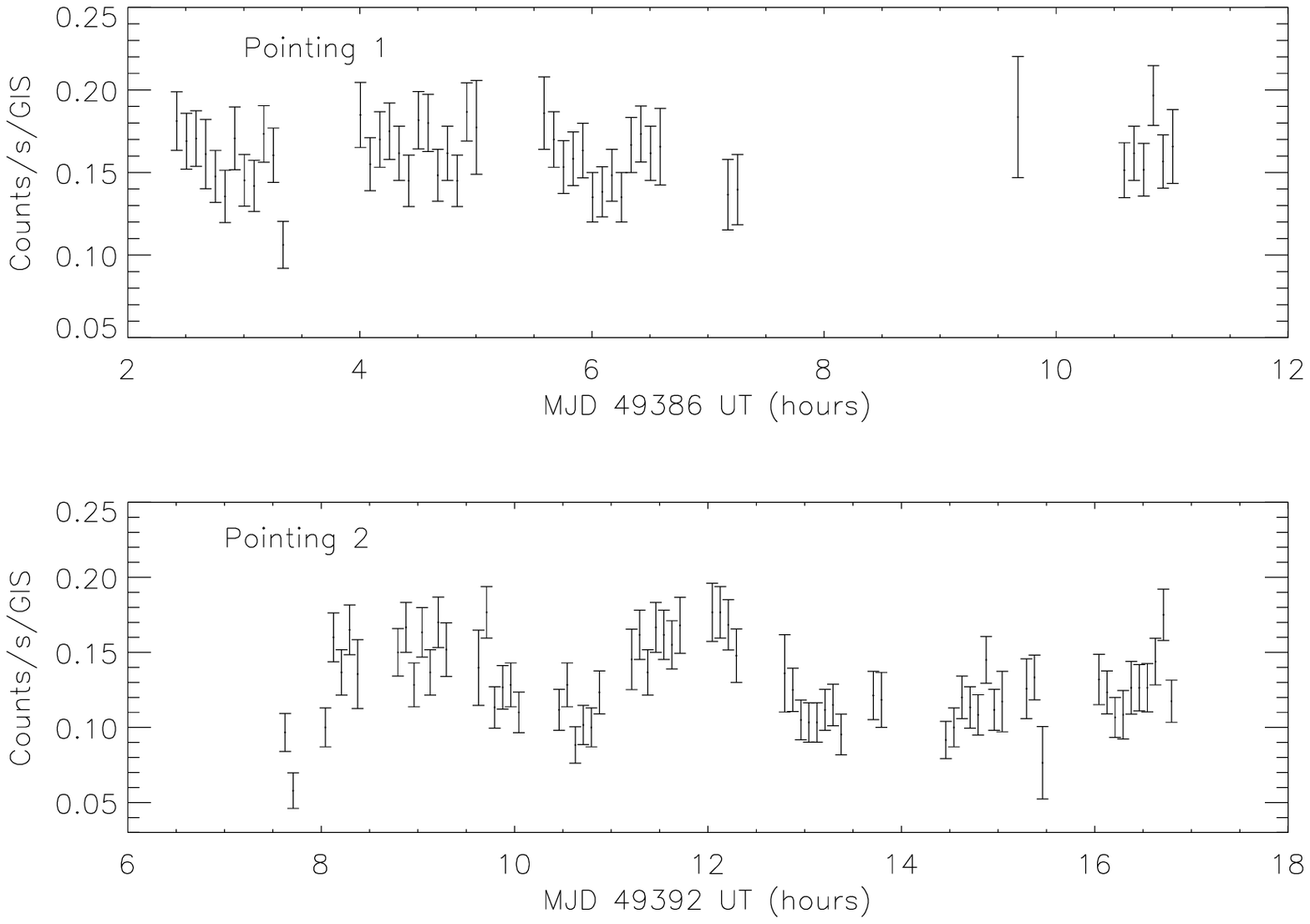,width=6.0in}}
\caption[lightcurve.eps]{Countrate in the 1 -- 5 keV band from \lsi\ for the two
\asca\ pointings.  The first pointing is consistent with no variability
on any timescale, whereas during the second pointing the source
varies by $\sim$50\% on timescales of 30 minutes. Error bars
are 1$\sigma$.\label{fig-lightcurve}}
\end{figure}

Small amplitude flares with timescales for rise and fall of $\sim$1.5
hours have been observed in the radio (\cite{per97}).  We searched for
similar variability in the archival data.  The ROSAT data were very
sparsely sampled, with the pointings being interrupted by long data
gaps.  It was therefore not possible to search for variability on hour
timescales.  Background variations on 90 minute orbital timescales in
the RXTE PCA data also preclude searching for this periodicity in these
data.  The two ASCA pointings were interrupted only by Earth
occultations and SAA passage, and we searched for variability in the
countrates in the two GIS detectors.  We found that during the first
pointing (P1) the flux is consistent with a constant value, however
during the second (P2) the X-ray emission is variable on similar
timescales to those seen in the radio, but with relatively large
amplitude. Figure~\ref{fig-lightcurve} shows lightcurves, binned in 5
minute intervals, from the two GIS detectors combined for the two
pointings.  The variability by more than 50\% is evident in the second
data set.  The short gaps in the lightcurves are due to passage
through the South Atlantic Anomaly and Earth occultations.

In order to search for smaller-amplitude variability during P1, and
characterize the fastest time scales of the variability seen during
P2, we used a Lomb normalized periodogram (Press {\em et al.}
1992\nocite{pt+92} and references therein). This technique does not
introduce features due to data gaps (such as SAA passages and Earth
occultations) which would normally appear using standard Fourier
analysis, since the data are evaluated only at times when they are
actually measured. A periodogram of the data from P1 indicates no
power at any frequency up to the Nyquist frequency, and is entirely
consistent with Poisson noise.  Highly-significant power is, however,
seen in the data from P2 with the highest frequency of
$10^{-4}~$s$^{-1}$, indicating variability on timescales of $\sim$40
minutes.  This implies a limit on the characteristic size for the
emission region of $\lsim$6$\times 10^{13}$ cm, or $\sim$4 AU.

\subsection{Periodicity Searches}

Several models for the behavior of this source invoke a rapidly
spinning, magnetized neutron star.  If such a star is present, and if
at least some of the X-ray emission observed originates within the
magnetosphere of this neutron star, then we expect to find modulation
of the X-ray flux at the neutron star rotation period.  We have
searched both the XTE and ASCA data for coherent pulsations.

\subsubsection{RXTE}

We searched each section of nearly continuous data (a total of 36
observations, typically 1000 -- 4000 seconds long) independently for
evidence of coherent periodicities.  For this search we used the
``GoodXenon'' data, which retain full time resolution (1 $\mu$s) and
energy information about every non-vetoed X-ray photon.  We selected
these data for good time intervals, events in the 2--10 keV range,
and in Layer 1 only. We then corrected event times to the Solar System
Barycenter so that Doppler shifts from the motion of the Earth and the
spacecraft would not smear out a coherent signal from the source.
We generated light curves with 1 ms resolution (1 to 4 $\times 10^6$
bins).

Because the neutron star is being accelerated in its orbit around the
companion B star, a search over orbital acceleration is required to
keep short period pulsars from being rendered undetectable due their
period changing during the observation.  Thus, we applied a range of
orbital accelerations to the time series before performing the
standard pulsar search.  This consists of doing a Fast Fourier
Transform, generating a normalized power spectrum, and searching for
narrow significant peaks both singly and in harmonically related
groups.  We found no significant peaks in the range 0.01--500 Hz in
any of the observations down to a pulsed fraction of $\sim$6\% for an
individual 1000 second data segment.

Finally, to improve the sensitivity of the search to slow pulsars ($>$
a few hundred ms) for which orbital acceleration is negligible, we
averaged all 36 unaccelerated power spectra and searched for peaks.
Again, no significant periodicities were found.  This search was
sensitive to a pulsed fraction of $\sim$1\%.  In both cases, pulsed
fractions are calculated for a sinusoidal modulation and the fraction
is the fraction of the {\em total} count rate (including source,
cosmic background, and instrumental background).

As a test of the XTE extraction software and of our pulsation search
code, we extracted data from the known rotation-powered X-ray pulsar
PSR 1509--58 which was taken during the XTE In Orbit Checkout period
and is part of the XTE public archive.  The GoodXenon data from a 2000
second observation of PSR 1509--58 was then subjected to the same
filtering, barycentering and processing as the data from \lsi.  The
6.627 Hz pulsation and at least three harmonics were easily detected at a
significance of 58$\sigma$.

\subsubsection{ASCA}

\begin{table}
\caption{ASCA periodicity search limits}
\label{tab-period}
\begin{center}
\begin{tabular}{|l|c|c||c|c|}
\hline
 & \multicolumn{2}{|c||}{P1} & \multicolumn{2}{|c|}{P2}\\ 
\hline
\lsi & \# events & Amplitude Limit & \# events & Amplitude Limit \\
\hline
GIS 1 & 2638 & 0.28 & 2159 & 0.24 \\
GIS 2 & 2629 & 0.28 & 2585 & 0.22 \\
GIS 1 + 2 & 5267 & 0.20 & 4744 & 0.16 \\
\hline
Source2 &  &   &   &  \\
\hline
GIS 1 & 455 & 0.68 &  690  & 0.42 \\
GIS 2 & 434  & 0.70 & 738  & 0.40 \\
GIS 1 + 2 & 889 & 0.49 & 1428 & 0.29 \\
\hline
\end{tabular}
\end{center}
\end{table}

In addition to the {\em RXTE} search, we performed a pulsation search
on the {\em ASCA} data.  For the GIS detectors during P1 the time
resolution of 0.0625~s limits the search to a minimum period of
0.125~s, and for P2, the minimum period searched is 1.0~s,
corresponding to the Nyquist limit for the 0.5~s sampling.  Epoch
folding using the standard XRONOS software (available from the High
Energy Astrophysics Science Archive Research Center (HEASARC) at
Goddard Space Flight Center) failed to detect periodicity in either
pointing.  Table~2 summarizes the 90\% confidence upper limits on the
amplitude of pulsed emission for searches on the separate and combined
GIS 1 and 2 data.  We calculated these limits using the formalism
described in Leahy {\em et al.}  (1983)~\nocite{lde+83}.  A second
source in the ASCA FOV, identified as a stellar source (labeled
Source2) was also searched for pulsations, and the upper limits are
also given in Table~2.  For P1, in addition to the epoch folding, we
performed an FFT, which yielded nearly identical upper limits on the
pulsed amplitude fraction as the epoch folding.  For P2, aperiodic
variability in the source precluded the use of an FFT for period
searching (see \S~\ref{sec-aperiodic}).  We also searched the SIS data
to a minimum period of 8 s, (determined by the 4 s CCD readout time),
and found no pulsations, with limits similar to those presented for
the GIS.

To search for short-period pulsations, for which smearing of the
pulsed signal due to variable Doppler shifts resulting from the binary
orbital motion can be important, we performed an acceleration search
on the GIS data from P1.  The variation in the pulse frequency is
well-approximated by a linear trend over the duration of each
observation.  We searched a range of constant frequency derivatives
from $-3.6 \times 10^{-7} s^{-2} < \dot{\nu} < 3.6 \times 10^{-7}
$s$^{-2}$ (consistent with the range expected from the orbital
solutions) in 1600 spacings. No periodicity was detected.

\section{Discussion}



Several common characteristics of the two simultaneous, well-sampled
X-ray and radio lightcurves (Figures \ref{fig-xtelight} and
\ref{fig-rosatlight}) are evident.  Both are clearly modulated at the
orbital period, and the peak in the X-ray emission is significantly
offset (by 0.4 -- 0.5 in phase) from the peak in the radio emission.
Although containing only two data points, the {\em ASCA} observations
are also consistent with the anticorrelation of radio and X-ray
emission.  In the {\em ROSAT} 0.5 -- 2 keV data, the ratio of peak to
quiescent luminosity is a factor $\sim$20, similar to what is observed
in the radio, while in the {\em RXTE/PCA} 2--10 keV data, the
variation is a factor 3 -- 7 (the range depending on the uncertain
zero point).  Because of the poor spectral resolution and narrow
bandpass of the {\em ROSAT} data, and the difficulty in determining
spectral parameters in the {\em RXTE/PCA} data, it is impossible to
determine if this difference arises from variable X-ray absorption
(potentially important in the soft {\em ROSAT} band but negligible for
the PCA), variation in the spectral index, or variation in the X-ray
outburst intensity orbit-to-orbit.

These observations suggest that it is unlikely that both the X-ray and
the radio flux arise from synchrotron emission from a common electron
population.  Rather, it is more likely that the X-ray flux results
from inverse Compton scattering of stellar photons off a quasi-steady
relativistic electron population.  First, it is difficult to
accomodate a significant offset between the X-ray and radio peaks if
both result from synchrotron emission from a common population.
Furthermore, for reasonable synchrotron models for the radio emission,
inverse Compton losses should dominate at X-ray wavelengths over the
entire orbit.  For example, Peracaula (1997)\nocite{per97} has modeled the
synchrotron emission from an expanding spherical plasmon ejected from
the binary system to explain the radio emission. To fit these data, the
initial radius is $r_o = 1.5 \times 10^{13}$~cm, the distance from the
star is the orbital separation ($a = 7. \times 10^{12}$ cm), the
magnetic field is $B = 0.15$~G, and the expansion velocity is $3
\times 10^{7}$~cm/s.  Electrons are injected uniformly into the
expanding sphere at a steady rate over 7 days.  In this model, to get
the radio rise requires a large initial $r_o$, and for the implied
magnetic field and $r_o$, inverse Compton losses will dominate over
synchrotron losses at all distances from the star.
For synchrotron losses to be important at X-ray
energies requires $B \gsim 300$~G for an electron population with
spectral index 2.1 (steeper spectral index requires a higher field).

The small-amplitude of the orbital modulation of the X-ray flux
relative to the radio observed by {\em RXTE} indicates that the
electron population responsible for the inverse Compton emission is
relatively constant, with the flux modulation likely resulting from
variations in the stellar photon density as a function of position in
an eccentric orbit. The X-ray maximum will occur near periastron,
where the stellar photon density is highest (although some offset is
possible if the Be star equator is not aligned with the orbit). The
factor 3 -- 6 X-ray variations would imply $r_{min}/r_{max} =
1/\sqrt{3} - 1/\sqrt{6}$, or orbital eccentricity $\epsilon = 0.27 -
0.42$ for a steady electron population.  This range is consistent with
the poorly-defined orbital parameters, which allow a range of
eccentricity from 0.2 -- 0.8~(\cite{hc81}).

As originally proposed by Marashi and Treves (1981)\nocite{mt81}, a
relatively steady electron population could be produced at the shock
resulting from the interaction of the relativistic wind of a young
pulsar with the Be stellar wind.  The pulsar wind would create a
cavity, with the boundary determined by dynamic pressure balance.
These electrons inverse Compton scatter the stellar photons, producing
X-ray emission, modulated by the variation in photon flux due to an
eccentric orbit.  The observed 2 -- 10~keV X-ray luminosity of $(1 - 6)
\times 10^{33} (D/(2.3 \kpc))^2$~erg~s$^{-1}$ is consistent with the
shock interpretation, and is similar to the observed luminosity near
periastron of PSR~B1259$-$63~(\cite{ktn+95}), for which a similar
mechanism has been invoked~(\cite{tak94}).  This model is also
consistent with the lack of observed X-ray pulsations, since the X-ray
emission is produced well outside the pulsar magnetosphere.

To explain the periodic radio
outbursts requires an expanding plasmon.  This could be produced when,
sometime after periastron, the stellar wind pressure at the back side
of the cavity becomes low enough that the cavity is no longer closed,
allowing relativistic particles to stream away from the system,
producing the radio emission and the observed expanding VLBI
source. In this scenario, the X-ray peak should occur at periastron,
and the radio outburst approximately 0.5 in phase away, as is
observed.  If 2CG 135+01 is associated with the system, the 100~MeV
gamma-rays could also be produced via the inverse Compton mechanism by
electrons of energy $\sim 3 \times 10^{-3}$~erg.  This emission should
also be modulated at the orbital period.  The EGRET flux from
2CG~135+01 is, in fact, time variable on day
timescales~(\cite{tkm+98}), however the statistics are not sufficient
to establish an orbital modulation.

\subsection{High-energy Spectrum}

\begin{figure}
\centerline{\psfig{file=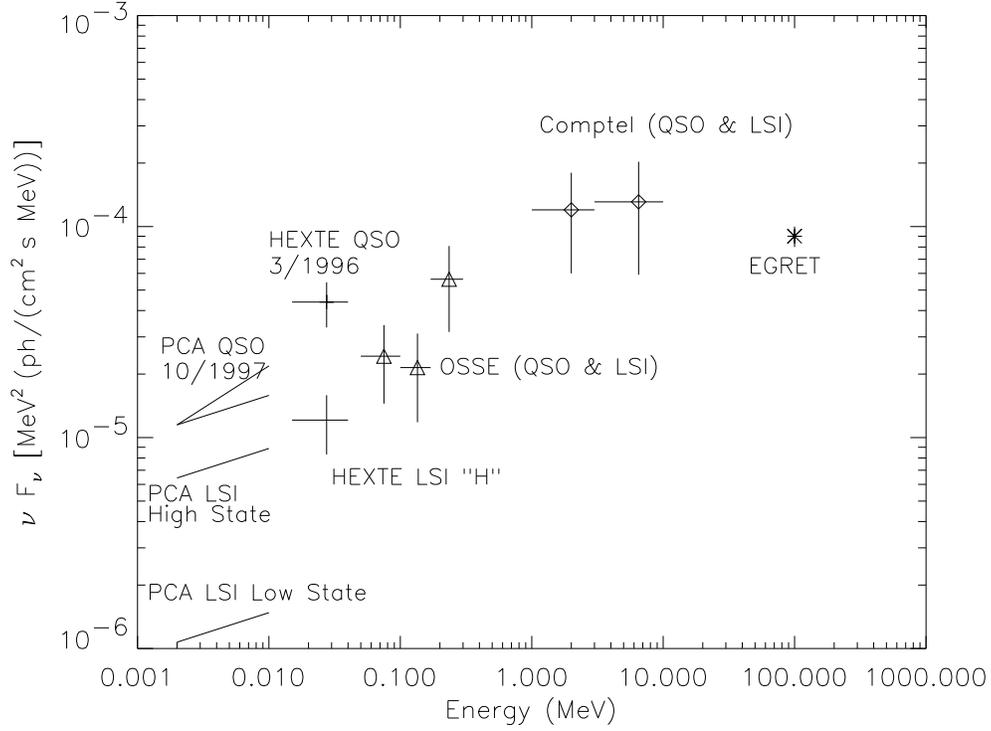,width=6.0in}}
\caption[fig-hespec.eps]
{
Compilation of selected high-energy observations of the region
containing \lsi\ and QSO~0241+662. Previous observations in the 100
keV -- MeV band by OSSE (triangles) (Strickman
{\em et al.}  1998) and COMPTEL (diamonds) (van Dijk et al. 1996)
included both \lsi\ and QSO~0241+662, while {\em
RXTE} resolved both sources in the 2--150 keV band.  
The range of 2--10 flux for
\lsi\ indicates the variation from high-state to low state, and
includes in the range uncertain background subtraction.  It is clear
that the OSSE and COMPTEL measurements are dominated by flux from
QSO~0241+662, since the observation periods are long, and average over
a large fraction of the \lsi\ orbit.  The EGRET observation (Thompson
{\em et al.} 1995) excludes emission from QSO~0241+662, but the
13\arcmin\ error radius makes association with \lsi\ uncertain.
\label{fig-hespec}}
\end{figure}

Numerous instruments have claimed detection of \lsi\ at hard X-ray and
gamma-ray energies, however few have had positioning capability
sufficient to distinguish between emission from the binary and
QSO~0241+622 (QSO~0241+622 being the only source within a several
degree radius with a known hard spectrum).  

Figure~\ref{fig-hespec} summarizes some of the most secure hard X-ray
and gamma-ray detections from this region.  As described previously,
the PCA and HEXTE clearly detect both sources.  In the PCA, the quasar
2 -- 10~keV flux is a factor $\sim$2 higher than the average
high-state of \lsi.  In HEXTE, the quasar 15 -- 40~keV flux
(determined at a different epoch than the PCA quasar measurement) is a
factor 2 -- 7 higher than the HEXTE \lsi\ detection.  This would tend
to indicate variability of the QSO, however the statistics are
limited.  To derive PCA fluxes for
\lsi , given the uncertainties in the PCA background subtraction and
response for the pointing, we fit a power law model with photon index
1.8 (consistent with the {\em ASCA} values), and we show a range that
includes uncertainty in the zero level.  For the HEXTE quasar
detection, we used the standard background subtraction, and fit a flat
spectrum over the 15 - 40 keV band to determine the flux. The OSSE and
COMPTEL measurements shown in the figure included both \lsi\ and
QSO~0241+622 in the instrument fields of view
(\cite{SSC+98,vdbb+96}).The EGRET angular resolution is sufficient to
exclude the quasar as the source of the gamma-ray emission (but the
13\arcmin\ error radius makes positive association of the 100 MeV flux
with \lsi\ uncertain).

It is clear from the {\em RXTE} observations that both the OSSE and
COMPTEL measurements were dominated by flux from the quasar.  Both
observations integrated over times comparable to the \lsi\ orbit, and
should be consistent with the average \lsi\ flux, which is a factor
$\sim 5$ lower than the QSO flux.  

\section{Conclusions}

Our observations have confirmed that the X-ray and soft gamma-ray
emission from \lsi\ is variable, and modulated on the orbital
timescale.  The {\em RXTE} observations sample a harder band than the
previous {\em ROSAT} data, and we can eliminate variable absorption as
the source of the flux modulation.  From the two well-sampled
lightcurves, it is clear that the X-ray emission peaks almost half an
orbit before the radio.  One explanation for this offset is that the
high-energy emission is produced by inverse Compton scattering of
stellar photons by a population of relativistic electrons produced at
a nearly steady rate at the shock interaction of the relativistic wind
of a young pulsar and the Be star wind.  This would produce an X-ray
peak near periastron. The synchrotron radio emission would then arise
from the expansion of the plasma near apastron, when the cavity
produced by the pulsar is no longer confined. Our observations fail to
detect X-ray pulsations that would signify the presence of a young
neutron star.

We also detected X-ray emission from the nearby QSO~0241+622 in the 2
-- 150 keV band.  The quasar is an approximate factor of 2 brighter
than the average high-state \lsi\ flux at these energies.  Previous
100 keV -- MeV observations of the field made by OSSE and COMPTEL were
therefore likely dominated by the quasar emission.  The broadband
1~keV -- 100~MeV spectrum of \lsi\ therefore remains uncertain.
Future high-energy missions, such as {\em GLAST}, will have the
angular resolution and flux sensitivity to both positionally associate
2CG 135+01 and
\lsi , and detect orbital modulation in the 100~MeV flux, if
it exists.

\acknowledgements

We would like to thank Keith Jahoda and the XTE GOF staff for
assistance with the XTE software and background models.  Portions of
this work were funded by NASA under the RXTE Guest Investigator
program.  Basic research in X-ray and Radio Astronomy at the Naval
Research Laboratory is supported by the Office of Naval Research.


\end{document}